\documentclass[useAMS,usenatbib]{mn2e}
\usepackage{graphicx}
\usepackage{natbib}
\usepackage{ amssymb}
\usepackage{color}
\title[Cross Correlations: The 21 cm and the NIRB]{Stars and Reionization: The Cross-Correlation of the 21 cm Line and the Near-Infrared Background}
\author[Elizabeth R. Fernandez, Saleem Zaroubi, Ilian T. Iliev, Garrelt Mellema, Vibor Jeli\'{c}]{Elizabeth R. Fernandez\thanks{Fernandez@astro.rug.nl}$^1$, Saleem Zaroubi$^1$, Ilian T. Iliev$^2$, \newauthor Garrelt Mellema$^3$, Vibor Jeli\'{c}$^{1,4}$\\
$^1$Kapteyn Astronomical Institute, The University of Groningen,  
Landleven 12,               
9747 AD Groningen,       
The Netherlands\\
$^2$Astronomy Centre, Department of Physics \& Astronomy, Pevensey II Building, University of Sussex, Falmer, Brighton BN1 9QH, UK\\
$^3$ Department of Astronomy \& Oskar Klein Centre, Stockholm University,
AlbaNova, SE-10691 Stockholm, Sweeden\\
$^4$ ASTRON - the Netherlands Institute for Radio Astronomy, PO Box 2, 7990 AA Dwingeloo, the Netherlands
}%
\begin{document}
\label{firstpage}
\maketitle
\begin{abstract}
With improving telescopes, it may now be possible to observe the Epoch of Reionization in multiple ways.  We examine two of these observables - the excess light in the near-infrared background that may be due to high redshift stars and ionized HII bubbles, and the 21 cm emission from neutral hydrogen.  Because these two forms of emission should result from different, mutually exclusive regions, an anticorrelation should exist between them.   We discuss the strengths of using cross-correlations between these observations to learn more about high redshift star formation and reionization history.  In particular, we create simulated maps of emission from both the near-infrared background and 21 cm emission.  We find that these observations are anticorrelated, with the strongest anticorrelation originating from times when the universe is half ionized.  This result is robust and does not depend on the properties of the stars themselves.  Rather, it depends on the ionization history.  Cross-correlations can provide redshift information, which the near-infrared background cannot provide alone.  In addition, cross-correlations can help separate foreground emission from the true high redshift component, making it possible to say with greater certainty that we are indeed witnessing the Epoch of Reionization.

\end{abstract}
\section{Introduction}
\label{sec:introduction}
Understanding high redshift stars and galaxies and their complex relationship with their environment is one of the goals of modern cosmology.  These early generations of stars and galaxies instigated one of the main phase-transitions that the Universe has undergone - reionization.  Therefore, understanding the history of the Epoch of Reionization (EoR) can lead to an understanding of the properties of these sources themselves, how they evolved, and how they reionized the Universe.   
Up until very recently, our knowledge of these sources and the EoR were limited to theory and simulations.  Lately, however, thanks to new powerful telescopes, 
we can hope to begin to observe the EoR itself.  

Diversifying observations of the EoR will become increasingly important in the coming years.  Each different observation will provide different constraints on the population of stars forming at high redshifts.  In this work, we will discuss two ways to observe the EoR.    
Each of these focus on different types of emitting regions and different physics, yet these observables are fundamentally linked.  
Neutral hydrogen will emit 21 cm emission, a result of the hyperfine transition in hydrogen.  
This emission will  
be the strongest at early times, when the Universe still contained a significant fraction of neutral gas.  
(For a review of 21 cm physics, see \citet{Furlanetto/etal:2006}).  
Since the 21 cm emission is line emission, mapping the emission line in frequency space will lead to a three-dimensional picture of the neutral intergalactic medium (IGM).  
Therefore, the study of the structure of 21 cm emission can lead to information on the sources responsible for reionization, the ionized bubbles surrounding them, and how reionization progresses with time (e.g. \citet{Madau/etal:1997, Tozzi/etal:2000, Furlanetto/etal:2004, loeb/zaldarriaga:2004, Wyithe/Loeb:2004, Kohler/etal:2005, Furlanetto/etal:2006, iliev/etal:2006, iliev/etal:2007, iliev/etal:2012, Mellema/etal:2006, Wyithe/Morales:2007, Datta/etal:2008, Lidz/etal:2008, Santos/etal:2008, Geil/Wyithe:2009, harker/etal:2010, Morales/etal:2010, Santos/etal:2010, mack/wyithe:2012, Zaroubi/etal:2012, Malloy/Lidz:2013}).     
Using the 21 cm line therefore gives an excellent probe for reionization history.  Emission from the 21 cm line will possibly soon be observed with the Low Frequency Array (LOFAR, \citet{ciardi/etal:2013, vanhaarlem/etal:2013, Yatawatta/etal:2013, zaroubi:2013}), as well as other telescopes, such as the Precision Array for Probing the Epoch of Reionization \citep{parsons/etal:2010}, the Murchison Widefield Array \citep{Tingay/etal:2013},  the 21 Centimetre Array \citep{wang/etal:2013}, and the Giant Metrewave Radio Telescope \citep{Pen/etal:2008}, as well as next generation arrays such as SKA \citep{Mellema/etal:2013} or an omniscope \citep{clesse/etal:2012}.

As time goes on, the ionizing radiation from stars will chip away more and more of this neutral hydrogen, creating ionized "bubbles" that will no longer be bright in 21 cm emission.  
However, these stars, galaxies, and HII regions will emit ultraviolet photons, which will be redshifted to be observed in the infrared.  
Many of these photons come from galaxies that are far below the detection limit of current high redshift galaxy surveys \citep{barkana/loeb:2000, salvaterra/ferrara:2006, wyithe/loeb:2006, kistler/etal:2009, Bouwens/etal:2010,Robertson/etal:2010,fernandez/shull:2011,munoz/loeb:2011}.  
Therefore, the cumulative light in the infrared is one of the few ways available to observe these small objects.  

These redshifted photons should make up a portion of the unresolved background in the infrared, particularly from 1 to 4 ${\rm{\mu}}$m.  
In addition, a host of models have been developed to describe both the intensity and fluctuation component of the near-infrared background (NIRB) due to high redshift stars  \citep{kashlinsky/etal:2002, kashlinsky/etal:2004,kashlinsky/etal:2005,kashlinsky/etal:2007,kashlinsky/etal:2012,santos/bromm/kamionkowski:2002, magliocchetti/salvaterra/ferrara:2003,salvaterra/ferrara:2003,
cooray/etal:2004,cooray/yoshida:2004,kashlinsky:2005,madau/silk:2005, fernandez/komatsu:2006, fernandez/etal:2010, fernandez/etal:2012fluc, fernandez/etal:2012, cooray/etal:2012,  fernandez/zaroubi:2013, yue/etal:2012}.  
These models have shown that it is possible that observations of the NIRB can reveal information about stars during the EoR, such as their mass and metallicity, as well as 
galactic mass, suppression history, and escape fraction.  
Therefore, these same stars that cause a decrease in the 21 cm emission will also cause an increase in cumulative light from their own emission, which will be  present in the redshifted infrared.  
Many have searched for evidence of the high redshift stars component of the NIRB with instruments such as
the Diffuse Infrared Background Experiment on COBE, the Infrared Array Camera on {\it{Spitzer}}, NICMOS on {\it{Hubble}}, {\it{AKARI}}, and will soon be observed with CIBER  \citep{dwek/arendt:1998,gorjian/wright/chary:2000,kashlinsky/odenwald:2000,wright/reese:2000,
cambresy/etal:2001,totani/etal:2001, wright:2001, kashlinsky/etal:2002,kashlinsky/etal:2004, kash/etal:2007, kashlinskyb/etal:2007c, kashlinsky/etal:2012, magliocchetti/salvaterra/ferrara:2003, odenwald/etal:2003,cooray/etal:2004, kashlinsky:2005, matsumoto/etal:2005}.

Our goal in this paper is to see to what extent the NIRB and 21 cm observations are anticorrelated.  Since the 21 cm emission should result from neutral regions and the NIRB emission should result from regions of star formation, emission should result from differing locations on the sky.  Combining these observations will give more information about these stars and galaxies.  The 21 cm line emission can give redshift information, something that it is very difficult for the NIRB alone to provide.  
 In addition, an anticorrelation will be indicative that the signal of both the NIRB and the 21 cm emission are from the EoR, since an anticorrelation is not expected in foreground measurements.

In \textsection \ref{sec:sims}, we discuss our simulation we use to generate our sky maps.  In \textsection \ref{sec:model} we discuss the models we use to create our simulated sky maps.  These line-of-sight maps are created using the procedure described in \textsection \ref{sec:los} and cross-correlated in \textsection \ref{sec:xcoor}.  We conclude in \textsection \ref{sec:conc}. 
We use the cosmological parameters $h=0.7$, $\Omega_{\rm{m}}=0.27$, $\Omega_{\rm{B}}=0.044$.  

\section{Simulations}
\label{sec:sims}
The work presented here is based on large-scale {\it{N}}-body and radiative transfer simulation of cosmic reionization
presented in \citet{iliev.etal:2013}, and based off the methodology in \citet{iliev/etal:2006}. This simulation utilized a volume of $425\,$Mpc$^{-1}$ per side. The 
structure formation {\it{N}}-body component followed $5488^3$ (165 billion) particles starting from redshift $z=300$, 
with particle mass of $3.7\times10^7\,M_\odot$ and a force softening scale of $3.87~$h$^{-1}$kpc comoving. This
provided the underlying density and velocity fields and the basis for calculating the ionizing source populations. 
The radiative transfer simulation used $504^3$ uniform-grid cells and provided the time evolution of the ionized 
patches. The more massive sources, with halo masses above $10^9\,M_\odot$, were based on the haloes directly resolved
in the {\it{N}}-body simulation (using a spherical-overdensity halo finder with density threshold of $178$ times the mean)
and are assumed to be unaffected by the photoheating of the gas, as their masses are above the Jeans mass for $10^4$~K 
photoheated gas. The lower-mass sources, with total masses of $10^8 M_\odot$ to $10^9 M_\odot$, are modelled based on a 
sub-grid extended Press-Schechter prescription based on the local density (Ahn et al., in prep). The star formation 
in such sources could be fully or partially suppressed during reionization. Our source model is discussed in detail
in \citet{iliev/etal:2006}.   

\begin{figure}
\centering \noindent
\includegraphics[width=9cm]{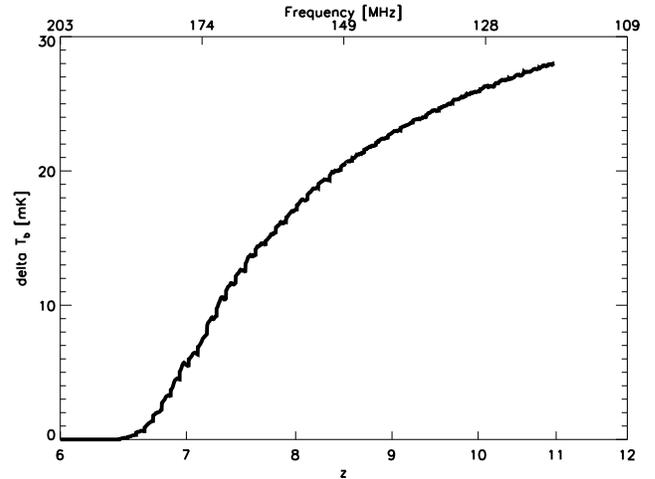}
\caption{ The differential brightness temperature ($T_{\rm{b}}$) of the 21 cm line from simulation, averaged over the simulation volume.}
\label{fig:tb}
\end{figure}

In order to be consistent with our underlying reionization model, we must describe how each source contributes
to reionization.  This is done by  
constraining the ionizing emissivity 
escaping into the IGM, which we assume to be proportional to the total halo mass.  This 
is quantified with a normalization parameter $f_\gamma$, which is a 
product of the fraction of baryons that form into stars ($f_*$), the escape fraction of ionizing radiation from the 
haloes into the IGM ($f_{esc}$), and the number of ionizing photons produced by stars per stellar 
atom ($N_i$) \citep{iliev/etal:2006}. This quantity, $f_\gamma = N_i f_{esc} f_*$, is set to a certain value in order 
to be consistent with our reionization scenario and the observational constraints, but each individual parameter is 
free to vary within this constraint. If, for example, there are more ionizing photons produced by a stellar population 
(if, for example, the stars are more massive and/or have less metals), either $f_{esc}$ or $f_*$ would correspondingly 
need to fall in order to avoid overproducing ionizing photons that escape into the IGM. In our radiative transfer 
simulation we adjust the value of $f_\gamma$ according to the mass of the haloes, where $f_{\gamma, {\rm{large}}} = 2$ for large 
haloes ($M> 10^9 M_\odot$), and $f_{\gamma, {\rm{small}}} = 8.2$ for small haloes ($10^8<M<10^9 M_\odot$).{\footnote{We can also express the normalization parameter independent of the simulation's star formation time-scale $t_{{\rm{sf}}}$ in terms of $g_\gamma=f_\gamma \frac{10 {\rm{Myr}}}{t_{{\rm{sf}}}}$, which is $g_\gamma=1.7$ for large haloes and $g_\gamma=7.1$ for small haloes.}  The value of $f_\gamma$ is consistent
with an electron-scattering optical depth of $\tau=0.0566$.  Reionization proceeds 
quite quickly, as illustrated by the ionization fraction as a function of redshift in Figure \ref{fig:ionfrac}, and 
completes by $z \sim 6.5$.

\begin{figure}
\centering \noindent
\includegraphics[width=9cm]{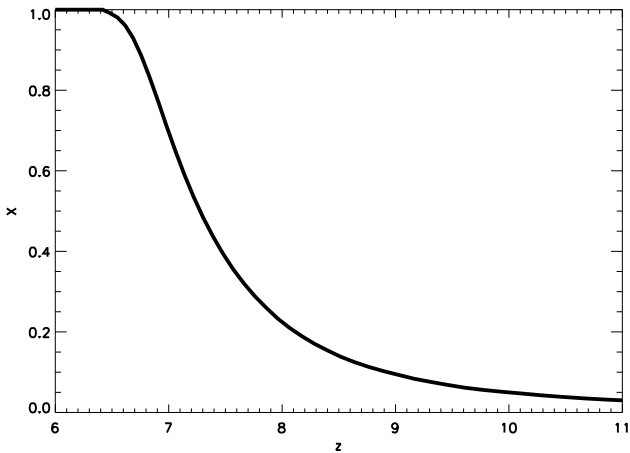}
\caption{ The ionization fraction by volume ({\it{X}}) of the IGM from the simulation as a function of redshift.  }
\label{fig:ionfrac}
\end{figure}

\section{Our Models}
\label{sec:model}
\subsection{The Near-Infrared Background}

The intensity of the NIRB will depend on many factors, namely, the properties of the stars and galaxies.  Emission that makes up the NIRB will originate from  the stars themselves.  Anything above 13.6 eV will be reprocessed as nebular emission, both inside and outside the halo.  
The main emission line we are concerned with is the Lyman-$\alpha$ line.  
There will also be significant continuum emission from other reprocessed nebular light (such as two-photon, free-free and free-bound emission; such as  \citet{fernandez/komatsu:2006}).  Because of this continuum, it is difficult to extract redshift information about the population of stars responsible for the NIRB (although not impossible, see \citet{fernandez/zaroubi:2013}).

In order to predict the emission of this high redshift population, some assumptions must be made about the stellar and galactic properties.  For our fiducial case, we assume that the stars are all Population II stars with a metallicity of $1/50 Z_\odot$.  The mass of these stars are assumed to follow a Salpeter mass function:
\begin{equation}
f(m) \propto m^{-2.35}
\label{eq:salpeter}
\end{equation}
\citep{salpeter:1955}, where $f(m)$ is the mass function, assuming lower and upper mass limits of $0.1$ and $150 \; \rm{M_\odot}$.     
We allow $10 \% $ of ionizing photons to escape into the IGM, $f_{esc}=0.1$.  $N_i$, the number of ionizing
photons produced per stellar atom, is calculated using fitting formulas in Table 3 of \citet{fernandez/komatsu:2008}, based of stellar models from \citet{marigo/etal:2001, lejeune/schaerer:2001} and \citet{schaerer:2002}.  Constraints from reionization, reflected in the value of $f_\gamma$, then set the fraction of baryons formed into stars to be $f_* =  7.7 \times 10^{-3}$ for large haloes and $3.2 \times 10^{-2}$ for small haloes.  Together,  these parameters create a consistent model for reionization.  
We relax the conditions of stellar mass, metallicity, $f_{esc}$ and $f_*$ in section \ref{sec:otherpops}.

After determining the relevant properties of the stellar population, there are two ways to calculate the luminosity per stellar mass ($l_\nu$) of each halo.  If the stellar lifetime is shorter than the star formation time-scale, dead stars must be taken into account:
\begin{equation}
 l^\alpha_\nu(z) 
= \frac{d\ln\rho_*(z)}{dt} 
\frac{\int^{m_2}_{m_1} {\rm{d}}m
   f(m)L^\alpha_{\nu}(m)\tau(m)}{\int^{m_2}_{m_1} {\rm{d}}m f(m) m}.
\end{equation}
The luminosity depends on the properties of the stars themselves - the stellar mass $m$, the luminosity $L_\nu(m)$ calculated for each component $\alpha$ (free-free, free-bound, two-photon, Lyman-$\alpha$ and stellar emission), the stellar lifetime $\tau(m)$, and the mass function.  
The stellar mass density, $\rho_*(z)$, can be written in terms of the star formation time-scale as $t_{sf}(z)=[d\ln\rho_*(z)/dt]^{-1}$.  Therefore, we can simplify the expression to:
\begin{equation}
 l^\alpha_\nu(z) = \frac1{t_{\rm SF}(z)}\frac{\int^{m_2}_{m_1} dm
   f(m)L^\alpha_{\nu}(m)\tau(m)}{\int^{m_2}_{m_1} dm f(m) m}.
\label{eq:lnu1}
\end{equation} 
For the simulation we use, the star formation time-scale is set at  $t_{sf} = 11.5$ Myr, which corresponds to  two time steps within our simulation.  

On the other hand, if we are concerned with small stars with long stellar lifetimes, it is possible that the stellar lifetime exceeds the star formation time-scale.  In this case, a different formula should be used to calculate the luminosity per mass: 
\begin{equation}
l^\alpha_\nu = \frac{\int^{m_2}_{m_1} dm f(m)
   L^\alpha_{\nu}(m)}{\int^{m_2}_{m_1} dm f(m) m}
\label{eq:lnu2}
\end{equation}
\citep{fernandez/etal:2010}.    
To be precise, a combination of Equations \ref{eq:lnu1} and \ref{eq:lnu2} should be used depending on the mass of the star \citep{ fernandez/komatsu:2006, fernandez/etal:2010}.  However, when we assume a Salpeter mass function of Population II stars, the luminosity-weighted mean lifetime of our stellar populations is longer than the star formation time-scale.  Therefore, to simplify the calculation, we use Equation \ref{eq:lnu2} as our fiducial formula.  In section \ref{sec:otherpops}, we consider other stellar populations, namely, one with massive Population III stars.  In this case, their luminosity-weighted mean lifetime is shorter than the star formation time-scale, so for these cases, we will use Equation \ref{eq:lnu1}.

The luminosity per mass ($l^\alpha_\nu$) of each of the relevant radiative processes 
is calculated following the formalism in \citet{fernandez/komatsu:2006}.  
To find the total luminosity of the halo, $L_{\rm{h}}$, the luminosity per mass $l_\nu^\alpha$ is multiplied by the baryonic mass of the halo that forms stars, ($M_{\rm{h}} f_* \frac{\Omega_{\rm{b}}}{\Omega_{\rm{m}}}$, where $M_h$ is the mass of the halo and $f_*$ is the fraction of baryons that form into stars),  and integrate over some bandwidth (so that $l^\alpha = \int^{\nu_2}_{\nu_1} l_\nu^\alpha d\nu$):
\begin{equation}
L_{{\rm{h}}}= M_{{\rm{h}}}\frac{\Omega_{{\rm{b}}}}{\Omega_{\rm{m}}} f_* [l^{*} + (1-f_{esc})(l^{{\rm{Ly}}\alpha}+l^{{\rm{ff}}}+l^{{\rm{fb}}}+l^{2\gamma})].
\end{equation}
The nebular components of the luminosity are multiplied by $(1-f_{esc})$, so that those photons that escape into the IGM do not contribute to the luminosity of the haloes.   These escaping photons will also contribute to the spectrum of the NIRB, but since the density of the IGM is low, the luminosity of the IGM will be much less than the haloes \citep{nakamoto/etal:2001, cooray/etal:2012}, and therefore, we ignore it for this work.

Finally, summing the luminosity over the haloes and dividing by the comoving volume, we can arrive at the volume emissivity in the band 
$p(z)$ (which describes the energy per unit time and unit comoving volume).  
The intensity is then found by integrating over redshift:
\begin{equation}
I = \frac{c}{4\pi} \int dz \frac{p(z)}{H(z){(1+z)}}
\end{equation}
\citep{peacock:1999}.  This intensity is a direct observable quantity that can be constrained from observations of the NIRB.

\subsection{The 21 cm Brightness Temperature}
Radiation will also result from the neutral IGM, manifesting itself as the 21 cm emission line.   
This is the result of a hyperfine transition in hydrogen.  
Since it results from neutral hydrogen, it is concentrated at high redshifts, before the majority of star formation occurred to ionize hydrogen.   Cold neutral gas can also be responsible for 21 cm absorption.  However, since we assume that the IGM is heated and $T_{\rm{S}} \gg T_{{\rm{CMB}}}$, here we take into account the emission only.

We calculate the differential brightness temperature of the 21 cm line ($\delta T_{\rm{b}}$):
\begin{eqnarray}
\delta T_{\rm{b}} &=& \frac{T_{\rm{S}}-T_{{\rm{CMB}}}}{1+z}(1-{\rm{e}}^{-\tau})\\ \nonumber
&\simeq& \frac{T_{\rm{S}}-T_{{\rm{CMB}}}}{1+z}\frac{3\lambda_0^3 A_{10} T_* n_{HI}(z)}{32 \pi T_{\rm{S}} H(z)}\\ \nonumber
&=& 27.0 \left[ \frac{1+z}{10} \right] ^{1/2} (1+\delta) \left( \frac{\Omega_{\rm{B}}}{0.044}  \frac{h}{0.7} \right) \left( \frac{0.27}{\Omega_{\rm{m}}} \right)^{1/2} 
\rm{mK}
\end{eqnarray}
\citep{field:1958, field:1959, Madau/etal:1997, ciardi/madau:2003, iliev/etal:2012}, 
where $T_{CMB}$ is the temperature of the cosmic microwave background (CMB), $T_{\rm{S}}$ is the spin temperature, $\tau$ is the optical depth, $\lambda_0 = 21.16$ cm is the rest frame wavelength of the 21-cm line, $A_{10}=2.85 \times 10^{-15} s^{-1}$ is the Einstein $A$-coefficient, $T_*$ is the energy difference between the two levels of neutral hydrogen, and $H(z)$ is the {\it{Hubble}} parameter.  
The value of $\delta T_{\rm{b}}$ depends on the overdensity of neutral hydrogen, written as,
\begin{equation}
1+ \delta = \frac{n_{HI}}{\langle n_{{\rm{H}}}\rangle },
\end{equation}
where $n_H$ is the total physical density of hydrogen and $n_{HI}=n_H(1-X)$ is the physical density of neutral hydrogen. 
This is a function of redshift, given by 
\begin{equation}
\langle n_{{\rm{H}}} \rangle (z) = \frac{\Omega_{\rm{b}} \rho_{crit,0}}{\mu_{\rm{H}} m_{\rm{p}}} (1+z)^3,
\end{equation}
where $m_p$ is the mass of the proton  and $\rho_{crit,0}$ is the present-day critical density.  We use $\mu_{\rm{H}} = 1.32$, which is the mean molecular weight of neutral hydrogen in units of the hydrogen abundance}, assuming a 24\% helium abundance.   
We show the brightness temperature as a function of redshift from the simulation in Figure \ref{fig:tb}, with the frequency of the observed 21 cm line shown on the top axis.  

\section{Line-of-Sight Maps}
\label{sec:los}

Combining the analytical models, described in Section \ref{sec:model}, with our simulation leads to three dimensional luminosity cubes at given redshifts.  
In order to move from these three dimensional cubes to the observed two dimensional sky, we must create line-of-sight maps.  
In order to create this line-of-sight, we rotate multiple simulation boxes randomly and stack them to create a cuboid, to allow the line-of-sight dimension to be longer than the simulation box size.   This rotation assures that structures along the line-of-sight are randomized and are not often repeated.  Multiple rotations allows us to use the same simulation data to create several randomizations for more statistically correct conclusions.  
This cuboid is then divided into slices, each corresponding to a redshift $z_k$.  
Since there may not be simulation output at $z_k$, the properties of the slice are determined by linear interpolation between the simulation outputs at 
two redshifts $z_i$ and $z_j$, such that $z_i<z_k<z_j$.   Here, redshift distortions are ignored.  We consider the effect of redshift distortions in Section \ref{sec:zdist}.

\begin{figure}
\centering \noindent
\includegraphics[width=8cm]{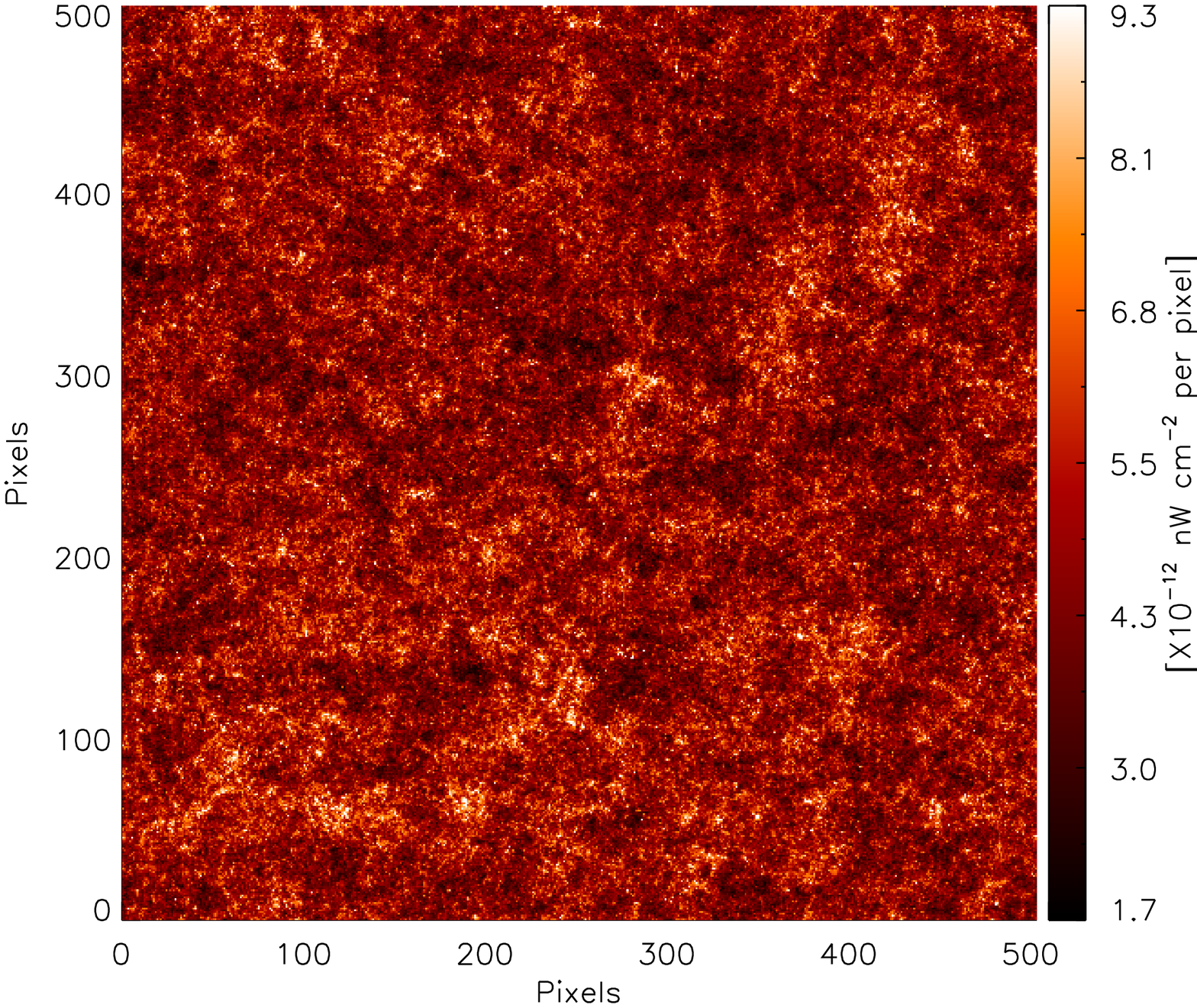}
\includegraphics[width=8cm]{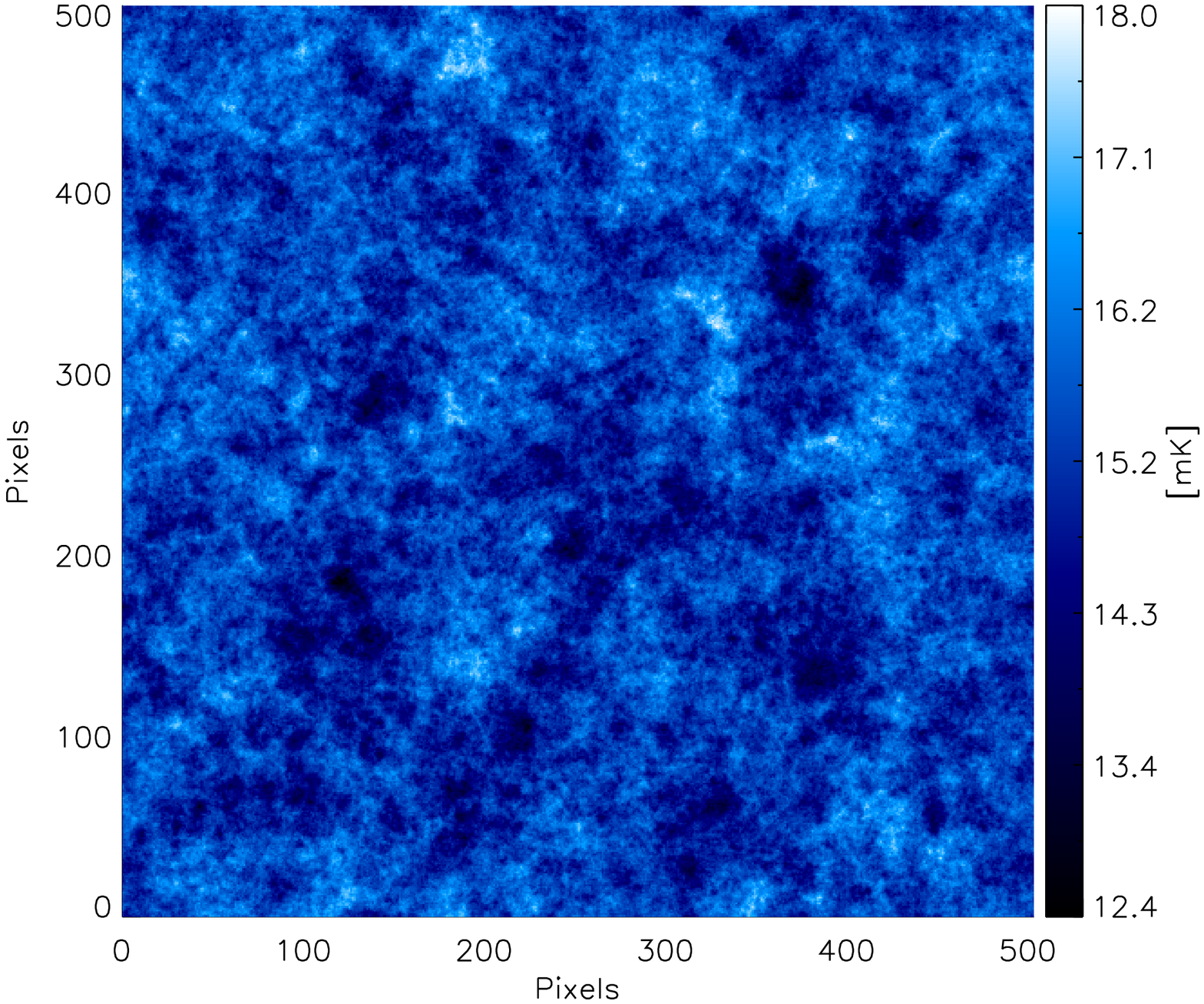}
\caption{The two-dimensional map from our simulation of the NIRB ({\it{top}}) and the 21 cm map ({\it{bottom}}) .  The field of view is about $3.75^{\circ}$ per side.  The map of the NIRB is in units of $\rm{nW} \rm{cm}^{-2}$ per pixel.  The LOFAR map is generated by averaging the sky map at each frequency within the LOFAR band, and is shown in mK.  
}
\label{fig:rawsky}
\end{figure}

A two dimensional observational map is then generated by projecting the information of each slice at $z_k$ onto the two dimensional sky.  
For the infrared sky, this is done by integrating these weighted maps over the redshifts we are concerned, $6<z<30$.  
We make the NIRB map in the $J$ band, from 1.1 to 1.4 ${\rm{\mu}}m$.  
For the 21 cm line emission, each frequency corresponds to a separate redshift, therefore we do not need to integrate over the line-of-sight.    
Instead, we stack all of the images at various frequencies for the cumulative map, shown only for illustration here.  In this paper, we focus on creating maps that mimic the observations from LOFAR.  However, similar maps can be constructed simulating the observations from other arrays designed to observe the 21 cm emission from the EoR, taking into account specific parameters such as the relevant frequency range, resolution, and noise levels.
For the LOFAR 21 cm maps,  
we set our maximum redshift to $z=11$, which corresponds to the highest redshift to which LOFAR is sensitive.  
Our lowest redshift is $z\sim 6.4$, which is the end of reionization within our simulation.  
 These maps are shown in Figure \ref{fig:rawsky}.   

Realistic LOFAR noise maps, assuming a three year integration time\footnote{Three year integration time corresponds to 3x600h=1800h.}, are then added to the 21 cm map from the simulation.
The noise
maps are generated using the standard LOFAR-EoR simulation pipeline.
[For details of the noise simulations, see e.g. \citet{chapman/etal:2012,chapman/etal:2013} and \citet{Zaroubi/etal:2012}.]  
These noise maps are consistent with the LOFAR instrument and have a pixel size of 70''.  The LOFAR maps have an $S/N<1$ per pixel, with an signal to noise ratio (S/N) of about 0.03 and 0.13 at 150 and 175 MHz respectively.  However, the S/N increases after smoothing, to 0.2 and 2.75 for 150 and 175 MHz respectively.  To add our simulated sky map to the LOFAR noise map, we smooth the simulated map over a Gaussian kernel with a full width at half-maximum of three pixels, which corresponds to the resolution of LOFAR.{\footnote {Noise is not added to our NIRB maps, since we do not consider only one observational set from a single telescope.}}   
This map is then added to the inverse variance weighted noise maps to create a prediction of the 21 cm emission plus noise.  
Finally, certain Fourier modes that LOFAR is not sensitive to are removed to make the final map, which is comparable to observations from LOFAR.

\section{The Cross-Correlation of the 21 cm and NIRB Maps}
\label{sec:xcoor}

We now have our two maps, one corresponding to the infrared sky, and the other corresponding to the 21 cm line emission.  
When looking at the total map of the NIRB, one can see the history of star formation during the EoR ($6<z<30$).  
Large scale structures can be seen across the sky, and bright areas correspond to areas of intense star formation, likely at lower redshifts.  
On the other hand, the 21 cm emission map shows areas where star formation is not occurring - thus, the bright areas are locations lacking star formation, with each frequency corresponding to a different redshift.  

\begin{figure*}
\centering \noindent
\includegraphics[width=8cm]{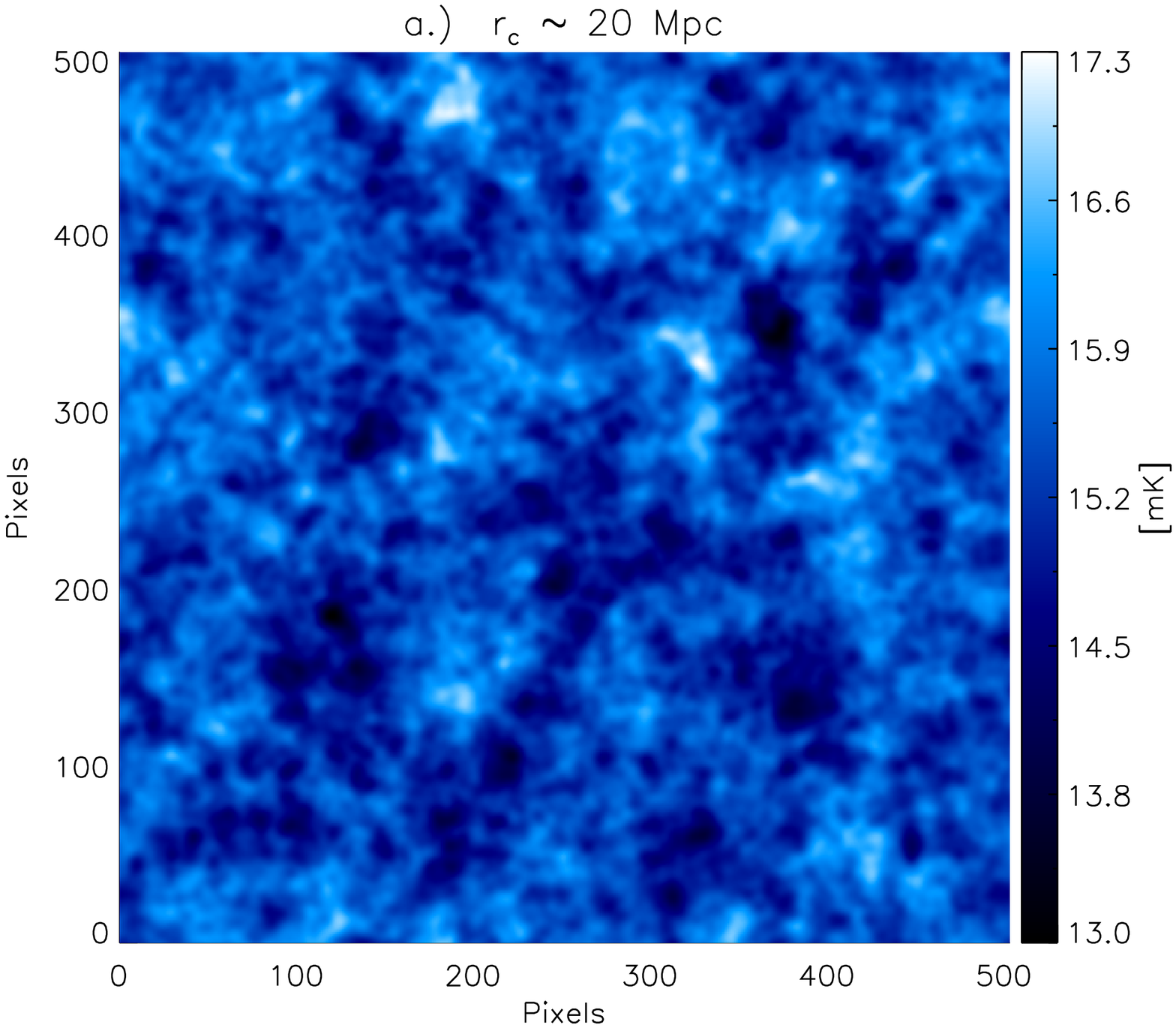}
\includegraphics[width=8cm]{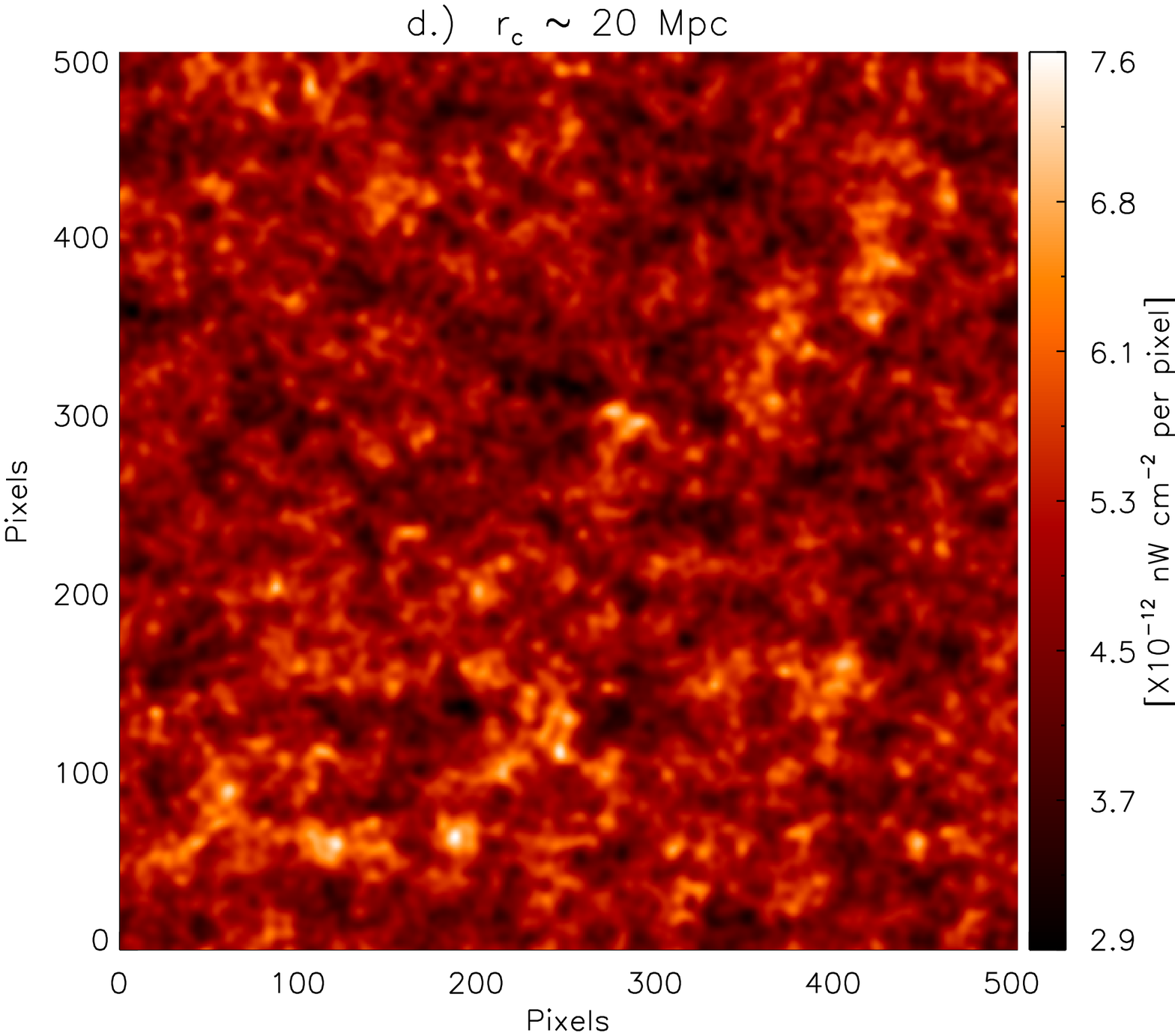}
\includegraphics[width=8cm]{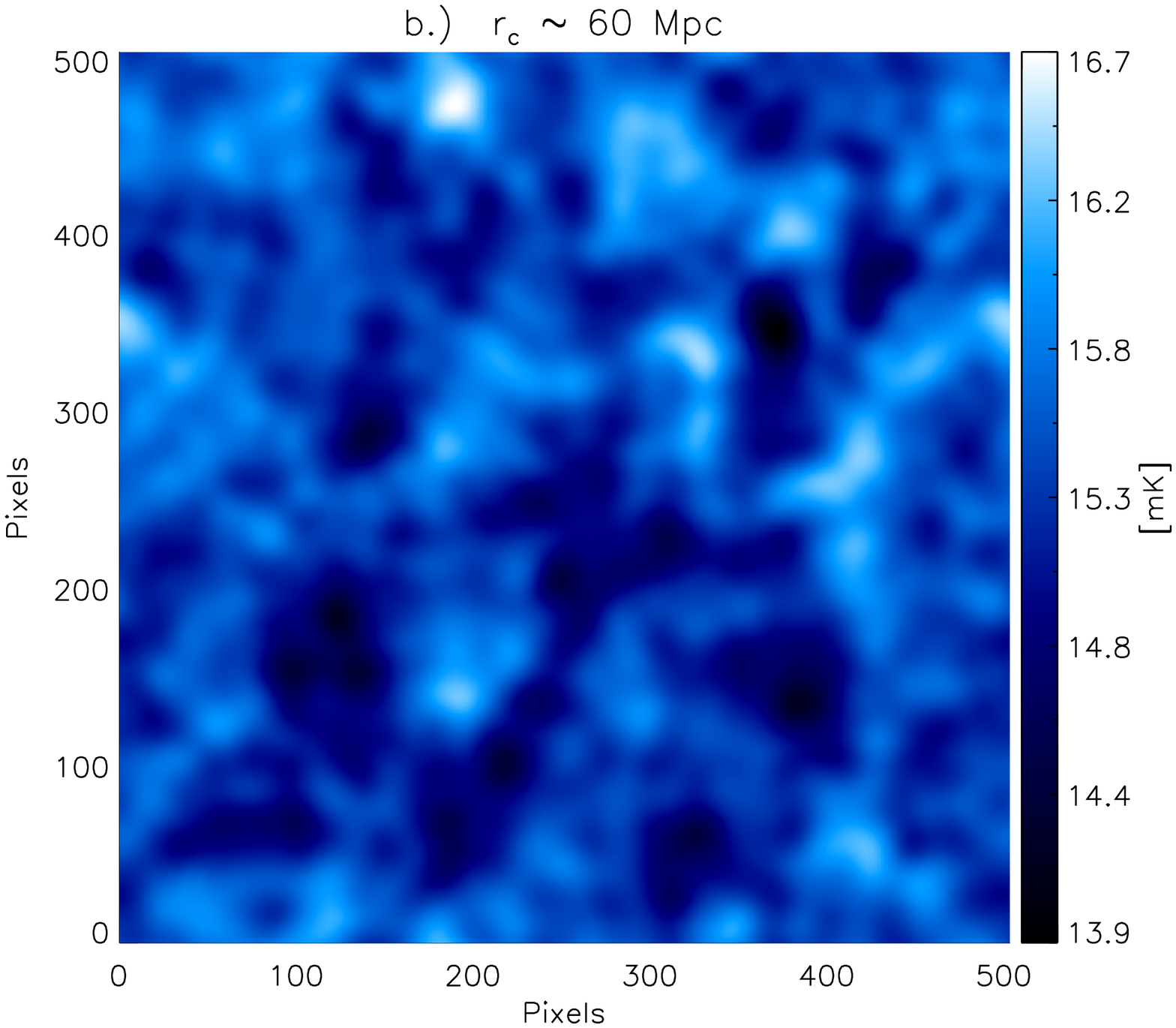}
\includegraphics[width=8cm]{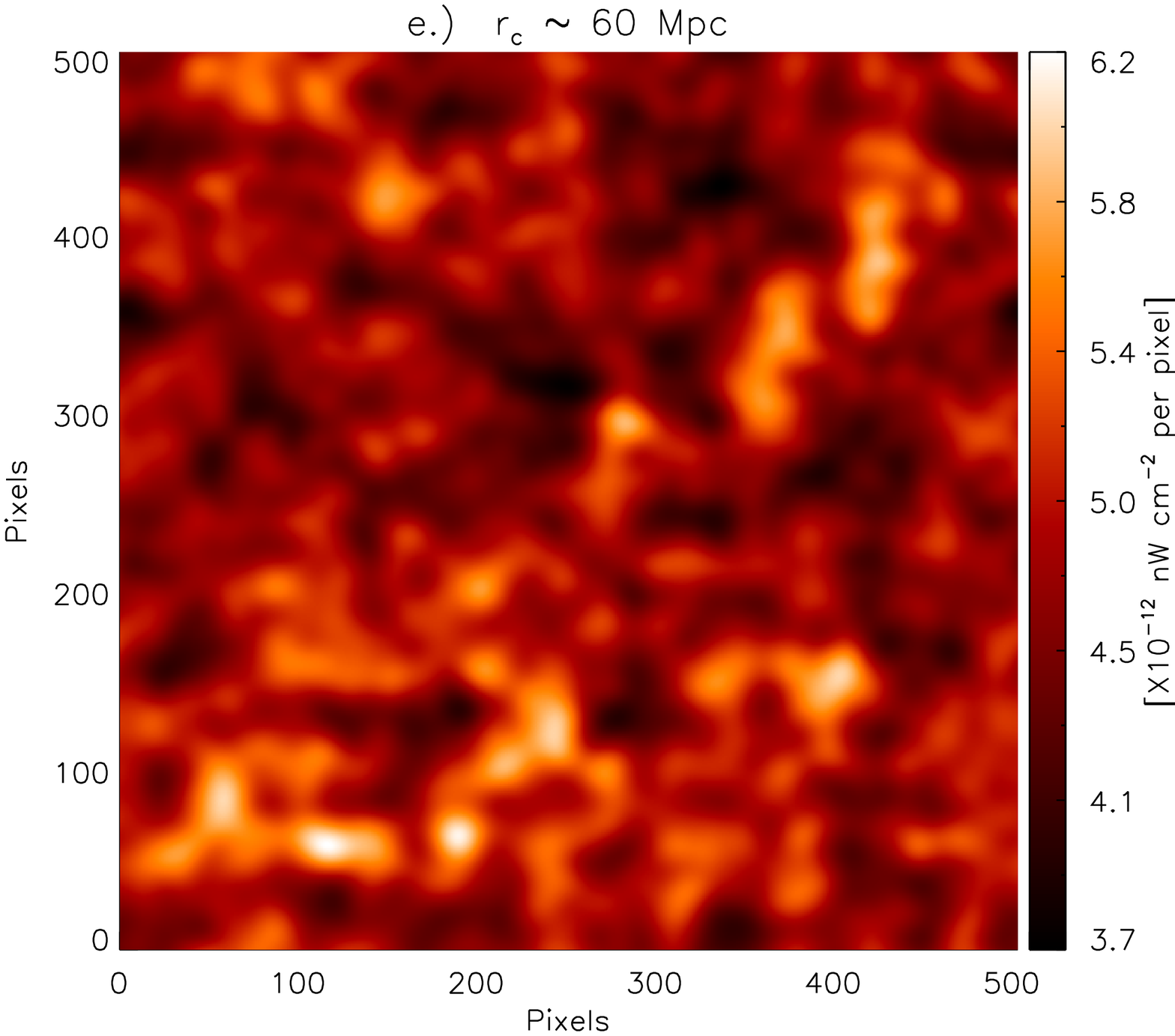}
\includegraphics[width=8cm]{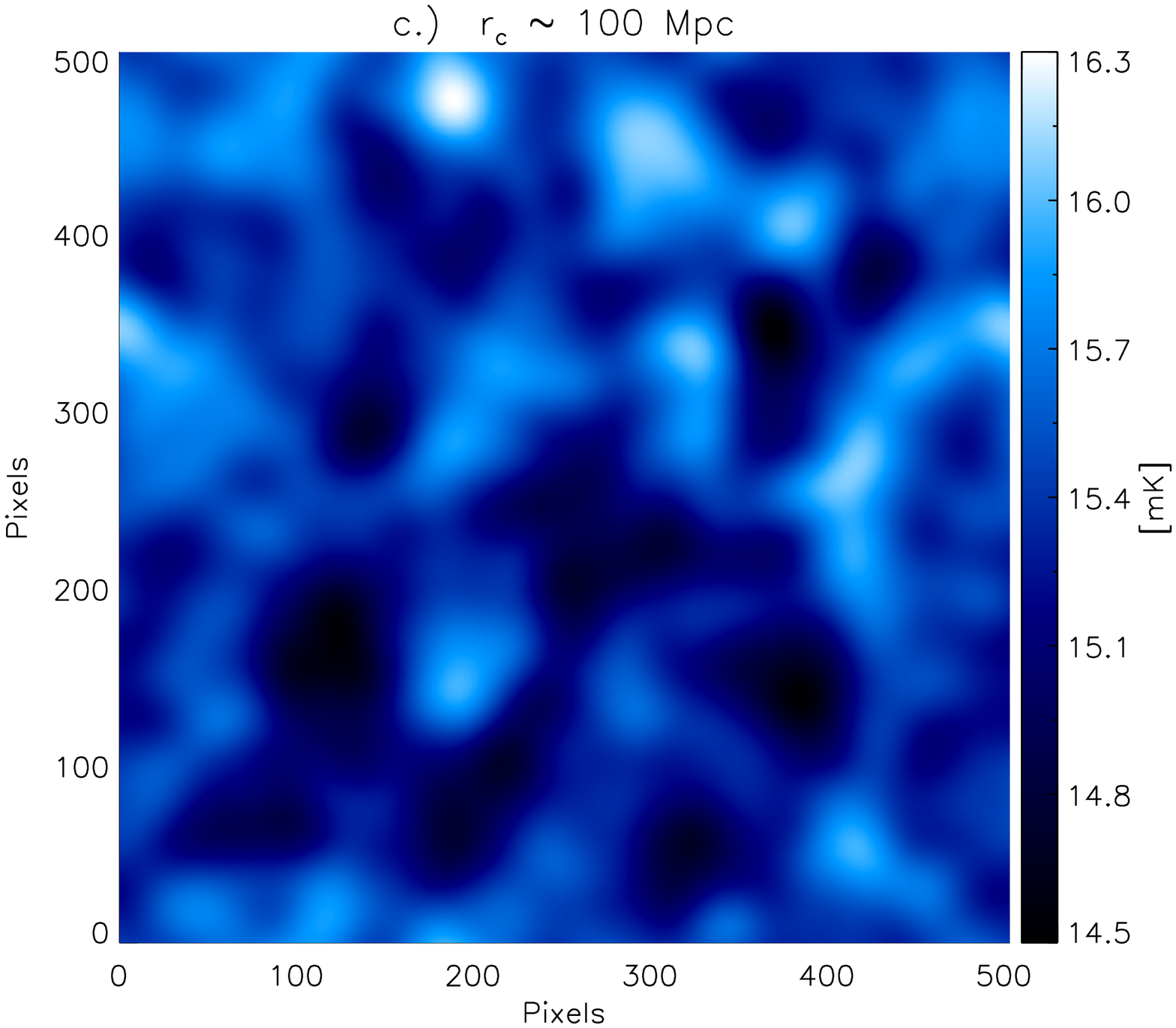}
\includegraphics[width=8cm]{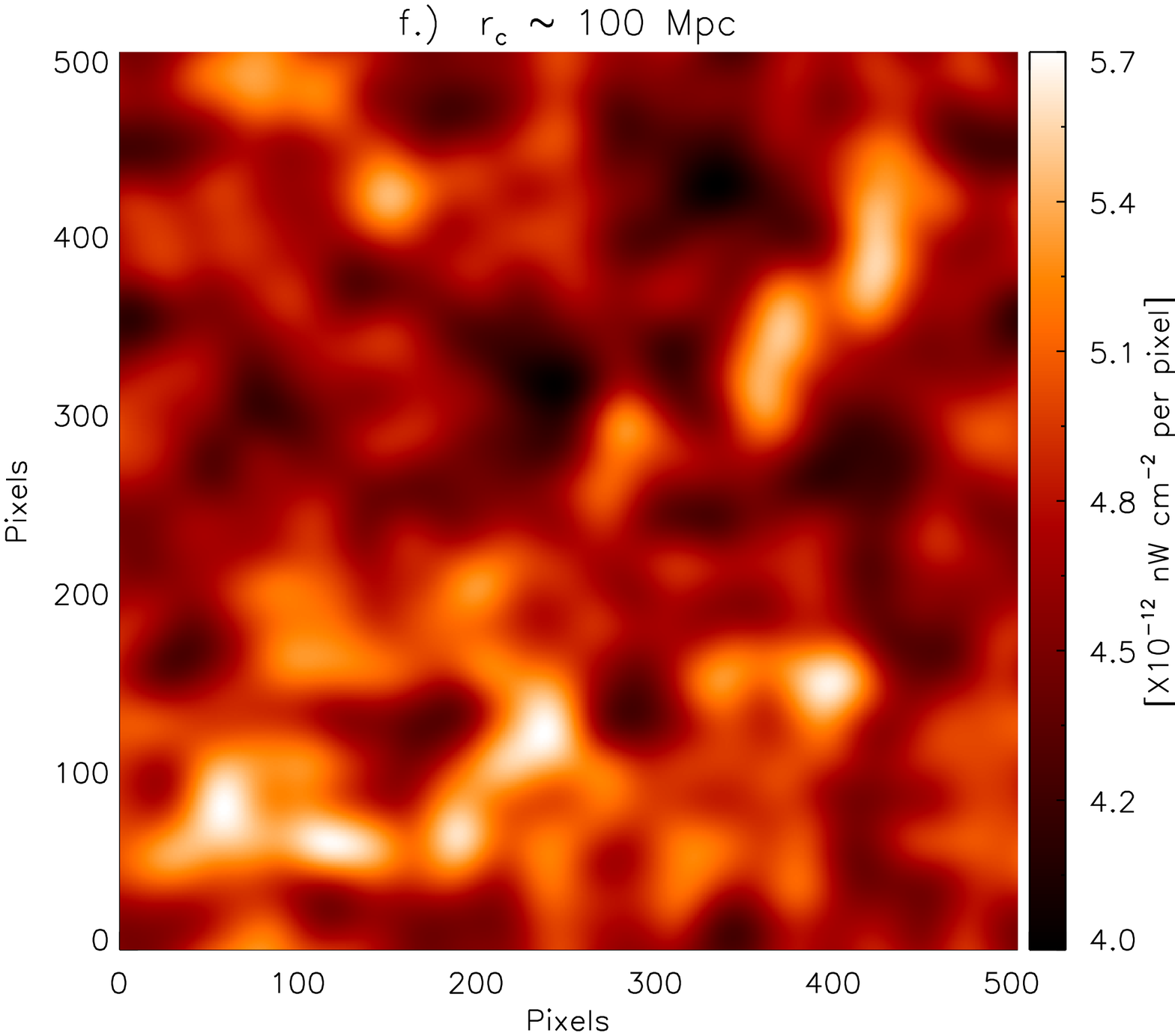}
\caption{The smoothed maps of the 21 cm emission ({\it{left, panels a through c}}) and NIRB ({\it{right, panels d through f}}) with various kernels, for $r_{\rm{c}}\sim 20$ Mpc ({\it{top row, panels a and d}}), $60$ Mpc ({\it{middle row, panels b and e}}), and $100$ Mpc ({\it{bottom row, panels c and f}}).    
Areas that are bright in the NIRB are dim in the 21 cm, and areas that are dim in the 21 cm are bright in the NIRB.  The map of the NIRB is in units of $\rm{nW} \rm{cm}^{-2}$ per pixel.  The LOFAR map is generated by averaging the sky map at each frequency within the LOFAR band, and is shown in mK.  The field of view is about $3.75^{\circ}$ per side.
}
\label{fig:smoothmap}
\end{figure*}

As is, the maps are too finely resolved to be directly cross-correlated with one another since the NIRB map is dominated by small-scale power.  Because of this, and also to more accurately represent the resolution of the LOFAR EoR experiment, we convolve the two dimensional map with a Gaussian over a smoothing radius $r_{\rm{c}}$.  Some examples of smoothed maps are shown in Figure \ref{fig:smoothmap} for various values of $r_{\rm{c}}$.{\footnote{Here, the noise on the LOFAR maps is not shown for illustration, since the noise overwhelms the 21 cm signal.  It is, however, included in the analysis.}}  While the image is convolved with a Gaussian kernel that corresponds to a certain number of pixels, the value of $r_{\rm{c}}$ shown in Figure \ref{fig:smoothmap} is the approximate size of the kernel in megaparsecs, calculated by taking into account the size of the image on the sky.

Even by eye, it is easy to see the anticorrelation between the two maps.  It is clear that areas that are bright in the 21 cm are in neutral areas that have not been ionized by star formation and therefore dim in the infrared.  
Areas that are bright in the NIRB have abundant star formation to ionize the regions around them, eliminating the possibility of a strong 21 cm line.

This cross-correlation between the intensity of the NIRB ($I_{NIRB}$) and the brightness temperature {$(\delta T_{\rm{b}})$} can be quantified by the Pearson correlation coefficient:
\begin{equation}
\rho_{21 {\rm{cm}}, {\rm{NIRB}}} = \frac{{\rm{cov}}({(\delta T_{\rm{b}}),{I_{NIRB}}})}{\sigma_{\delta T_{\rm{b}}}\sigma_{I_{NIRB}}}.
\end{equation}

During our cross-correlation analysis, the NIRB maps we use are always for the entire redshift range, $6<z<30$.  
Because the 21 cm is line emission, we are free to choose any redshift or frequency range.  
For our first case, we show the correlation between all of the frequencies available from the LOFAR maps to the entire NIRB map.  This means we are comparing star formation present in the NIRB map from $6<z<30$ to neutral hydrogen in the 21 cm map at $6.4<z<11$.  The redshift range is not exactly the same, yet since the NIRB map will be dominated by structure at the later stages of reionization, this provides the best opportunity to correlate the same structures in each map.  
In the following analysis, the cross-correlations were obtained by averaging the cross-correlation coefficient over five different box rotation realizations.  

\begin{figure}
\centering \noindent
\includegraphics[width=9.5cm]{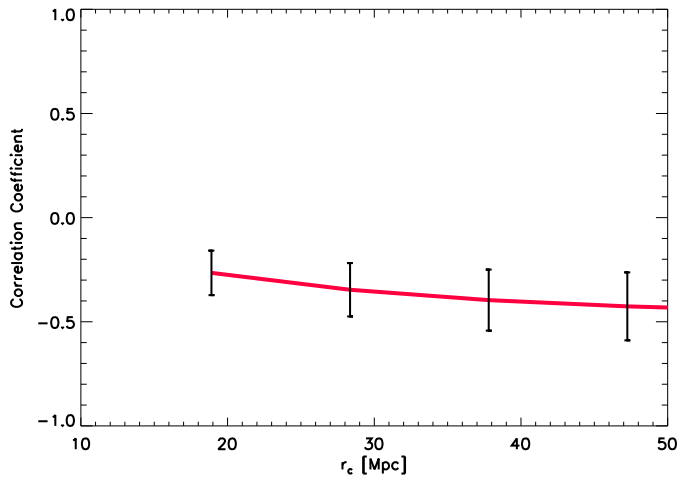}
\caption{The cross-correlation of the NIRB (from $6<z<30$) with the 21 cm signal (from $6.4<z<11$).  The error bars are generated by cross-correlating the 21 cm signal with 1000 realistic realizations of randomized NIRB maps.
}
\label{fig:corr}
\end{figure}

The value of this cross-correlation coefficient is shown in Figure \ref{fig:corr} for various values of the smoothing radius $r_{\rm{c}}$.  
The two emission maps are strongly anticorrelated, where -1 is a perfect anticorrelation, 0 is no correlation and 1 is complete positive correlation.  The error bars are determined from cross-correlating the 21 cm map with 1000 realistic realizations of a randomized map of the NIRB.  This is also in agreement with \citet{slosar/etal:2007}, who analytically found an anticorrelation of the of the cross-correlation power spectrum on small scales between the 21 cm and the cosmic infrared background.

Because LOFAR is sensitive to 21 cm line emission, emission at different frequencies will correspond to different redshifts.  
Therefore, we are free to only use observations at certain frequencies and correlate the entire NIRB to a limited redshift range of the 21 cm map.  
Therefore, we cross-correlate various narrow frequency maps of the 21 cm emission from our simulation independently against the entirety of the NIRB signal.  We expect the cross-correlation to weaken in these instances, since we are no longer comparing the emission of the NIRB and the 21 cm at similar ranges of redshifts, but rather comparing all of the NIRB emission with only a narrow range of the 21 cm emission.  

\begin{figure}
\centering \noindent
\includegraphics[width=9.5cm]{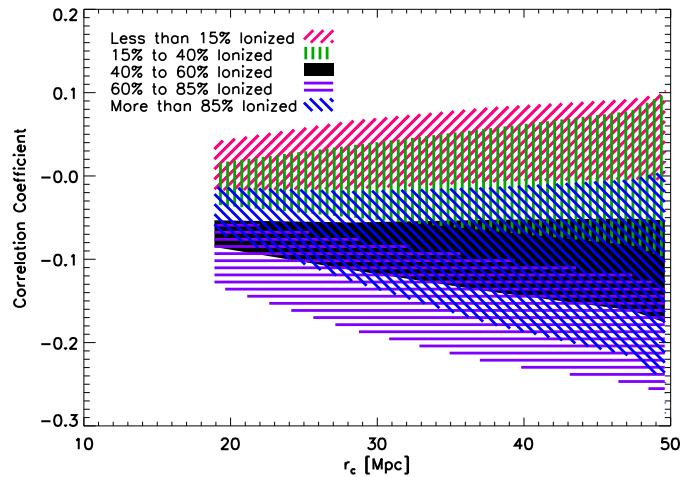}
\caption{The cross-correlation of the entire NIRB map against individual redshift slices of the 21 cm background maps.  The shaded regions represent the range of values that the correlation coefficient falls between when the ionization fraction of the particular 21 cm map falls between the given values.
}
\label{fig:corrall}
\end{figure}

This cross-correlation is shown in Figure \ref{fig:corrall}.  Each region corresponds to the range of values of the correlation coefficient falls between for various redshift slices of the 21 cm map.  We see that the strongest anticorrelation results when the universe is 50\% ionized or more, corresponding to mid to late reionization times.   At these times, the most structure is seen in the 21 cm maps.  At very early times, the anticorrelation becomes less pronounced, sometimes even becoming positively correlated.  This is because at early stages of reionization, the ionized bubbles are very small, and may be smaller than the smoothing length.  Therefore, both observations of the NIRB and the 21 cm emission pick up the high density peaks, rather than discriminating between ionized and neutral areas. 

\begin{figure}
\centering \noindent
\includegraphics[width=9.5cm]{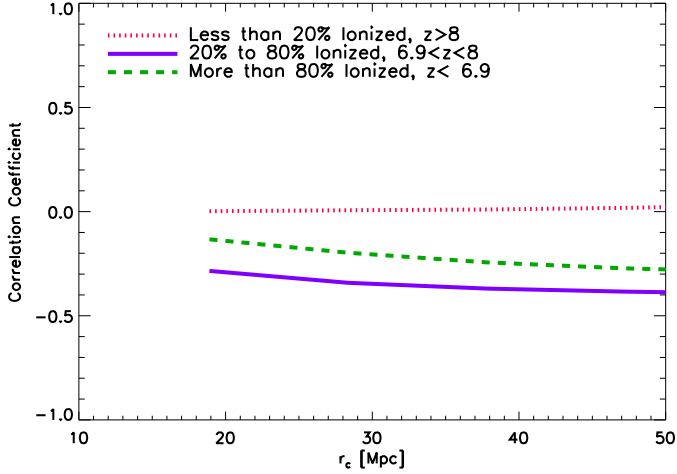}
\caption{The correlation coefficient of the entire NIRB against three cubes of the 21 cm emission - when the universe is less than 20\% ionized (early times, $z\gtrsim 8$), when the universe is more than 80\% ionized (late times, $z\lesssim 6.8$), and mid-reionization.  The strongest anticorrelation results when the universe is partially ionized. 
}
\label{fig:corrz}
\end{figure}

We can also selectively choose to combine certain ranges of frequencies of the 21 cm signal to correlate with the NIRB signal. In Figure \ref{fig:corrz}, we show the correlation coefficient when the 21 cm maps from only certain redshift intervals are cross-correlated with the NIRB.  Here, we can see that the strongest anticorrelation results from when the ionized fraction is around 20-80\%, which corresponds to the peak patchiness of the 21 cm fluctuations.

\subsection{Effects of the Stellar Population}
\label{sec:otherpops}
So far, we have assumed that all of the stars are Population II stars with a Salpeter mass spectrum and $f_{esc} = 0.1$.  We have also done all calculations assuming that our NIRB image is made in the $J$ band, from 1.1 to 1.4 ${\rm{\mu m}}$.  What were to happen if we were to vary these assumptions?  Because we do not know when stars transfer from metal-free to metal poor stars, we make an alternate assumption - that all stars are Population III, metal free stars.  We allow these stars to be massive, following the Larson mass spectrum \citep{larson:1998}:
\begin{equation}
  f(m) \propto m^{-1}\left(1+\frac{m}{m_c}\right)^{-1.35},
  \label{eq:larson}
\end{equation}
with mass limits of $0.1 M_\odot$ and $m_2=500 M_\odot$, and $m_c=250  M_\odot$.  In addition, we increase the escape fraction to $f_{esc}=1$, and, in order to be consistent with reionization history, we decrease $f_*$. 

We also modelled a case with a degree of metallicity evolution.  In this case, the large haloes within the simulation form Population II stars with a Salpeter mass spectrum, and the small haloes form Population III stars with a Larson mass spectrum.  This is the scenario that might be expected if the large haloes had a previous generation of star formation and are already metal enriched, while the smaller haloes are more pristine.  For this case, we set $f_{esc}=0.1$ for the large haloes and $f_{esc}=1$ for the small haloes.  

Finally, we calculated the cross-correlation when the band of NIRB map is changed to the $M$ band (from 4.6 to 5 ${\rm{\mu m}}$).  For this last case, we also assumed a massive Population III mass spectrum  (Equation \ref{eq:larson}) to provide the largest difference from our fiducial case.

\begin{figure}
\centering \noindent
\includegraphics[width=9.5cm]{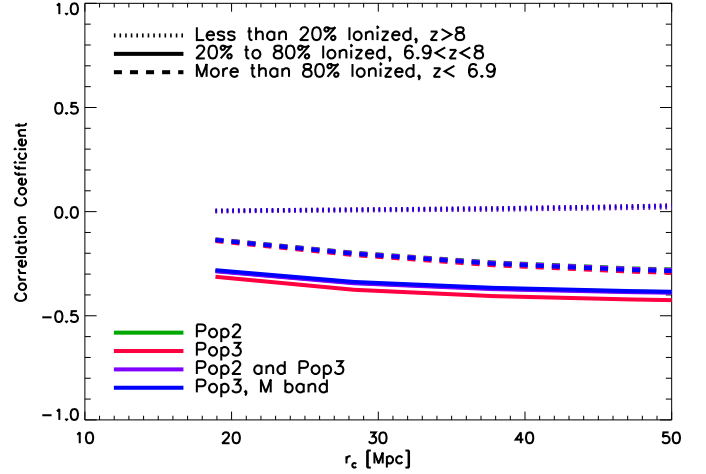}
\caption{The correlation coefficient of the entire NIRB and the 21 cm emission for various stellar populations.  Population II is our fiducial model, Population III assumes massive stars with zero metallicity, and Pop II/Pop III assumes Population II stars with a Salpeter mass spectrum in large haloes and Population III massive stars in smaller haloes.  We also tested a case where the NIRB observations were made in the $M$ band, while allowing the stars to be massive and metal free.  A significant change to the cross-correlations is not present.
}
\label{fig:coorpop}
\end{figure}

There is little effect in the overall cross-correlation when the stellar population is changed, as seen in Figure \ref{fig:coorpop}, even when we look at a completely different band.  Changing the properties of the stars, along with the band, will not affect the locations of star formation, just the intensity.  This intensity will not affect the overall cross-correlation.  

\subsection{Redshift Distortions}
\label{sec:zdist}

The 21 cm emission provides a way to directly link frequency with redshift.  However, because of random velocities, it is possible that distortions can be introduced into our observations, causing emission to appear at a different redshift than it is truly located.  
Redshift distortions come in two varieties.  
The Kaiser effect \citep{kaiser:1987} enhances clustering, and is a result of large scale coherent motion.  
It causes a compression along the line-of-sight.  
The Finger-of-God effect suppresses clustering, and is an effect of small-scale random motions.  
This causes an elongation of emission along the line-of-sight.   

To determine the extent of this effect, we also created a map of the 21 cm emission with redshift distortions included.  The observed redshift $z_{obs}$, affected by the velocity field, is related to the actual redshift $z$ by
\begin{equation}
z_{obs}= -1 + (1 + z) \sqrt{\frac{1 + v_\shortparallel/c}{1 - v_\shortparallel/c}},
\end{equation}
where $v_\shortparallel$ is the component of the velocity along the line-of-sight.  The redshift distortions actually distort each radiative transfer cell, stretching them or contracting them according to the velocity.  The signal is then remapped back to the distorted cell, sometimes decreasing or increasing the amplitude.  

The redshift distortions will only affect the 21 cm line emission, since changing the redshift will change the frequency at which the 21 cm emission appears \citep{Mao/etal:2012, Jensen/etal:2013, Shapiro/etal:2013}.  Since the NIRB is an integrated signal over a continuum, the effect of the redshift distortions will be minimal, and hence is ignored here.  

\begin{figure}
\centering \noindent
\includegraphics[width=9.5cm]{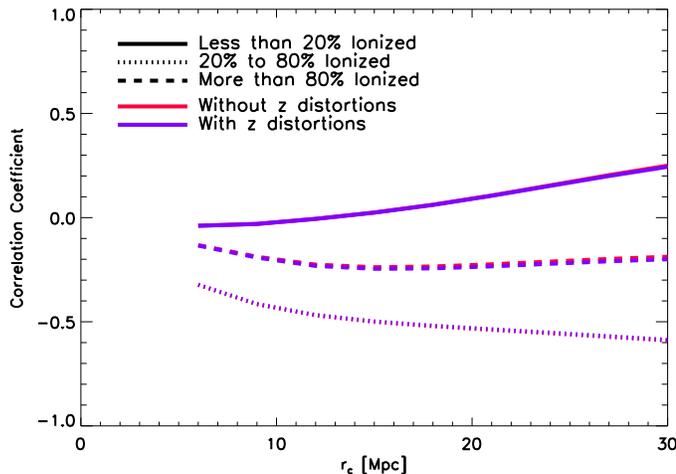}
\caption{The effect of redshift distortions on the results on one simulation realization, with and without redshift distortions, correlating the entire NIRB with redshift intervals of the 21 cm line.  While redshift distortions will change the cross-correlation strength for a narrow redshift range, once averaged over a larger slice, the effect disappears. Therefore, as long as the redshift range is not small, redshift distortions do not affect the result.
}
\label{fig:zdist}
\end{figure}

The results of adding redshift distortions on the cross-correlations is shown in Figure \ref{fig:zdist}  for one line-of-sight realization.  When a very narrow redshift range is correlated with the NIRB, a very small difference in the cross-correlation power is noticed,  as the brightness temperature is redistributed among cells.  However, when the redshift range is increased, this effect averages out, and adding redshift distortions has no effect on the cross-correlations.

\section{Conclusions}
\label{sec:conc}
With new generations of telescopes,  we hope to observe the EoR in multiple ways.  Combining these observations may lead to a deeper understanding of these high redshift stars and galaxies that would not be available from one observational set alone.  In this paper, we examined the cross-correlation between the 21 cm line, resulting from neutral hydrogen, and the high redshift component of the NIRB, resulting from areas of star formation.  

Because these two observations map out opposing regions, we would expect there to be an anticorrelation between these two observations.  In order to predict this, we generated simulated sky maps for both observations using our simulation data.  These maps show that the anticorrelation between these two observations exists and is quite strong.  The anticorrelation is the strongest when the ionization fraction is $\sim 50\%$.  There are many free parameters that can change the intensity of the high redshift component of the NIRB, such as the mass and metallicity of the stars, the escape fraction, the star formation rate, and the band of the NIRB observations.  However, these parameters do not change the cross-correlation.   
The cross-correlation probes {\it{only}} the reionization history, such as the redshift of reionization (where reionization is half completed) and the duration of reionization, and therefore, is an excellent method to examine how reionization progresses.  

Cross-correlations can also reduce some of the weaknesses these observations have on their own.  For example, it is very difficult to extract redshift information from the NIRB, but, by combining it with 21 cm line emission, more redshift information may be obtained.

Both observations of the NIRB and the 21 cm emission are quite challenging, namely due to a large amount of foregrounds that are difficult to model.   Incorrectly subtracted foregrounds would reduce the cross-correlation, so if a strong anticorrelation exists between the NIRB and the 21 cm emission, this would be one indication that the signal is from the EoR, since a similar anticorrelation should not exist when comparing foregrounds.  Therefore, if we observe a cross-correlation consistent with reionization history, we can say with greater certainty that we are indeed observing the EoR.

\section{Acknowledgments}
We acknowledge  the Dutch Science organization (NWO)
Vernieuwingsimpuls Vici programme for support.  This work was supported by
the Science and Technology Facilities Council [grant numbers ST/F002858/1
and ST/I000976/1]; and The Southeast Physics Network (SEPNet). We also
 acknowledge the Texas Advanced Computing Center (TACC) at The
University of Texas at Austin for providing HPC resources that have
contributed to the research results reported within this paper. This research
was supported in part by an allocation of advanced computing resources (PI
P. R. Shapiro) provided by the National Science Foundation through TACC and
the National Institute for Computational Sciences (NICS), with part of the
computations performed on Lonestar at TACC (http://www.tacc.utexas.edu)
and Kraken at NICS (http://www.nics.tennessee.edu/). Some of the
numerical computations were done on the Apollo cluster at The University of
Sussex and the Sciama High Performance Compute (HPC) cluster which is
supported by the ICG, SEPNet and the University of Portsmouth. Part of the
computations were performed on the GPC supercomputer at the SciNet HPC
Consortium (time granted to U.-L. Pen). SciNet is funded by: the Canada
Foundation for Innovation under the auspices of Compute Canada; the
Government of Ontario; Ontario Research Fund - Research Excellence; and
the University of Toronto.  GM is supported by Swedish Research Council
grant 2012-4144.
VJ is funded for this work by the Netherlands Foundation for Scientific
Research, VENI grant 639.041.336.

 \newcommand{\noop}[1]{}

\end{document}